%% file: combine.tex
\documentclass[
superscriptaddress,
reprint,
amsmath,
floatfix,
]{revtex4-1}

\usepackage{mathtools,amssymb,lipsum}
\usepackage{booktabs}
\usepackage{siunitx}
\usepackage[]{hyperref}
\usepackage{multirow}
\usepackage{bbold} 
\usepackage{float}
\usepackage[acronym]{glossaries}
\usepackage[normalem]{ulem}
\usepackage{csvsimple}
\usepackage{physics}
\usepackage{xcolor}
\usepackage{printlen}
\usepackage{tabularx}
\usepackage{tikz}
\usepackage{subfiles}

\usetikzlibrary{arrows}
\usetikzlibrary{decorations.pathmorphing}

\tikzset{snake it/.style={decorate, decoration=snake}}

\input{ion_macros} 

\newcommand{\thetitle}{A phonon laser in the quantum regime}

\newcommand{\theauthors}{
\setcounter{affil}{0}
\author{T. Behrle}
\email[E-mail: ]{tbehrle@phys.ethz.ch}
\affiliation{Institute for Quantum Electronics, ETH Z\"urich, Otto-Stern-Weg 1, 8093 Z\"urich, Switzerland}
\author{T.L.  Nguyen}
\altaffiliation{Current address: QuantX Labs Pty Ltd, Adelaide, Australia}
\affiliation{Institute for Quantum Electronics, ETH Z\"urich, Otto-Stern-Weg 1, 8093 Z\"urich, Switzerland}
\author{F. Reiter}
\affiliation{Institute for Quantum Electronics, ETH Z\"urich, Otto-Stern-Weg 1, 8093 Z\"urich, Switzerland}
\affiliation{Harvard University, 17 Oxford Street, Cambridge, MA 02138, USA}
\author{D. Baur}
\affiliation{Institute for Quantum Electronics, ETH Z\"urich, Otto-Stern-Weg 1, 8093 Z\"urich, Switzerland}
\author{B. de Neeve}
\affiliation{Institute for Quantum Electronics, ETH Z\"urich, Otto-Stern-Weg 1, 8093 Z\"urich, Switzerland}
\author{M. Stadler}
\affiliation{Institute for Quantum Electronics, ETH Z\"urich, Otto-Stern-Weg 1, 8093 Z\"urich, Switzerland}
\author{M. Marinelli}
\altaffiliation{Current address: JILA, Boulder, Colorado, USA}
\affiliation{Institute for Quantum Electronics, ETH Z\"urich, Otto-Stern-Weg 1, 8093 Z\"urich, Switzerland}
\author{F. Lancellotti}
\affiliation{Institute for Quantum Electronics, ETH Z\"urich, Otto-Stern-Weg 1, 8093 Z\"urich, Switzerland}
\author{S. F. Yelin}
\affiliation{Harvard University, 17 Oxford Street, Cambridge, MA 02138, USA}
\author{J. P. Home}
\email[E-mail: ]{jhome@phys.ethz.ch}
\affiliation{Institute for Quantum Electronics, ETH Z\"urich, Otto-Stern-Weg 1, 8093 Z\"urich, Switzerland}
\affiliation{Quantum Center, ETH Z{\"u}rich, 8093 Z{\"u}rich, Switzerland}
}

\begin{document}

\subfile{main}
\bibliography{./bibliography}
\newpage
\clearpage
\pagebreak

\subfile{supplemental_material}

\end{document}

%% file: ion_macros.tex
\newcommand{\create}{\ensuremath{{\,\hat{a}^{\dagger}}} }
\newcommand{\destroy}{\ensuremath{\,\hat{a}} }

%% file: main.tex
\title{\thetitle}
\theauthors

\begin{abstract}
We demonstrate a trapped-ion system with two competing dissipation channels, implemented independently on two ion species co-trapped in a Paul trap.  By controlling coherent spin-oscillator couplings and optical pumping rates we explore the phase diagram of this system, which exhibits a regime analogous to that of a (phonon) laser but operates close to the quantum ground state with an average phonon number of $\bar{n}<10$. 
We demonstrate phase locking of the oscillator to an additional resonant drive, and also observe the phase diffusion of the resulting state under dissipation by reconstructing the quantum state from a measurement of the characteristic function.  
\end{abstract}

\maketitle

Dissipation is intrinsic to quantum systems. While generally being viewed as an obstacle to implementing quantum control, it has also been shown to offer a resource for quantum state preparation and even quantum error correction \cite{Cirac1993, Poyatos1996,Carvalho2001, Verstraete2009, Diehl2008, Kraus2008, Pastawski2011, Krauter2011, Barreiro2011, Lin2013, Shankar2013, Sarlette2011, Navarette2014, Kronwald2013, Jarlaud, Kienzler2015, Malinowski2022, deNeeve2022, Kimich, Lu, Zapusek}. In the context of many-body physics, dissipation has been shown to lead to phase transitions many of which do not occur in fully coherent systems 
\cite{Diehl2008, Kessler2012, Morrison, Lang, Lee, Genway, Carmichael, Hannukainen, Hwang, Brennecke2013, Buchhold, Werner, Reiter2020, oetztuerk}. 
These aspects make the further investigation of dissipative dynamics in a range of physical systems of diverse interest. However open systems are challenging to deal with theoretically: numerical methods require simulating the evolution of a super-operator, which acts in a higher dimensional space than a unitary on the same underlying system. This makes it interesting to explore quantum simulation devices in which dissipation can be tuned precisely to explore all aspects of the physics 
\cite{Fink2018, Fink2017, Couedo2016, Ma2019, Fitzpatrick, Rodriguez, Bibak, GGE}. 
Realizing such dissipative simulators requires precise control of the coupling between the quantum degrees of freedom and the dissipation channels. 
When multiple dissipation channels are used, the broad-band nature of relaxation processes also produces stringent demands on crosstalk~\cite{Negnevitsky2018}. 

One archetypal example of a system in which dissipation plays a key role is the laser, which exhibits a dissipative phase transition from a dark phase into a bright phase characterized by the emergence of a coherent state with a random initial phase which diffuses over time. The parameters governing this transition are the photon loss rate (from the cavity mirrors) and gain due to the interaction with the gain medium, which saturates. While in general lasers operate with a gain medium featuring a large number of pumped systems \cite{loudon, scully_zubairy}, lasers have also been built at the single qubit level, e.g. using a single natural \cite{McKeever2003} or artificial atom \cite{Astafiev2007}. Analogous physics has also been observed using non-linear effects in the mechanical motion of objects spanning a range from atoms to nano-mechanics, e.g. realizing a ``phonon-laser'' \cite{Vahala2009, Grudinin2010, Kemiktarak2014, Cohen2015, Pettit2019,Wen2020, Zhang2021} with the lasing phase exhibiting large classical oscillations of tens of microns in size, corresponding to a mean value of $10^4$ phonons. 

\begin{figure}
    \centering
    \includegraphics[width=0.47\textwidth]{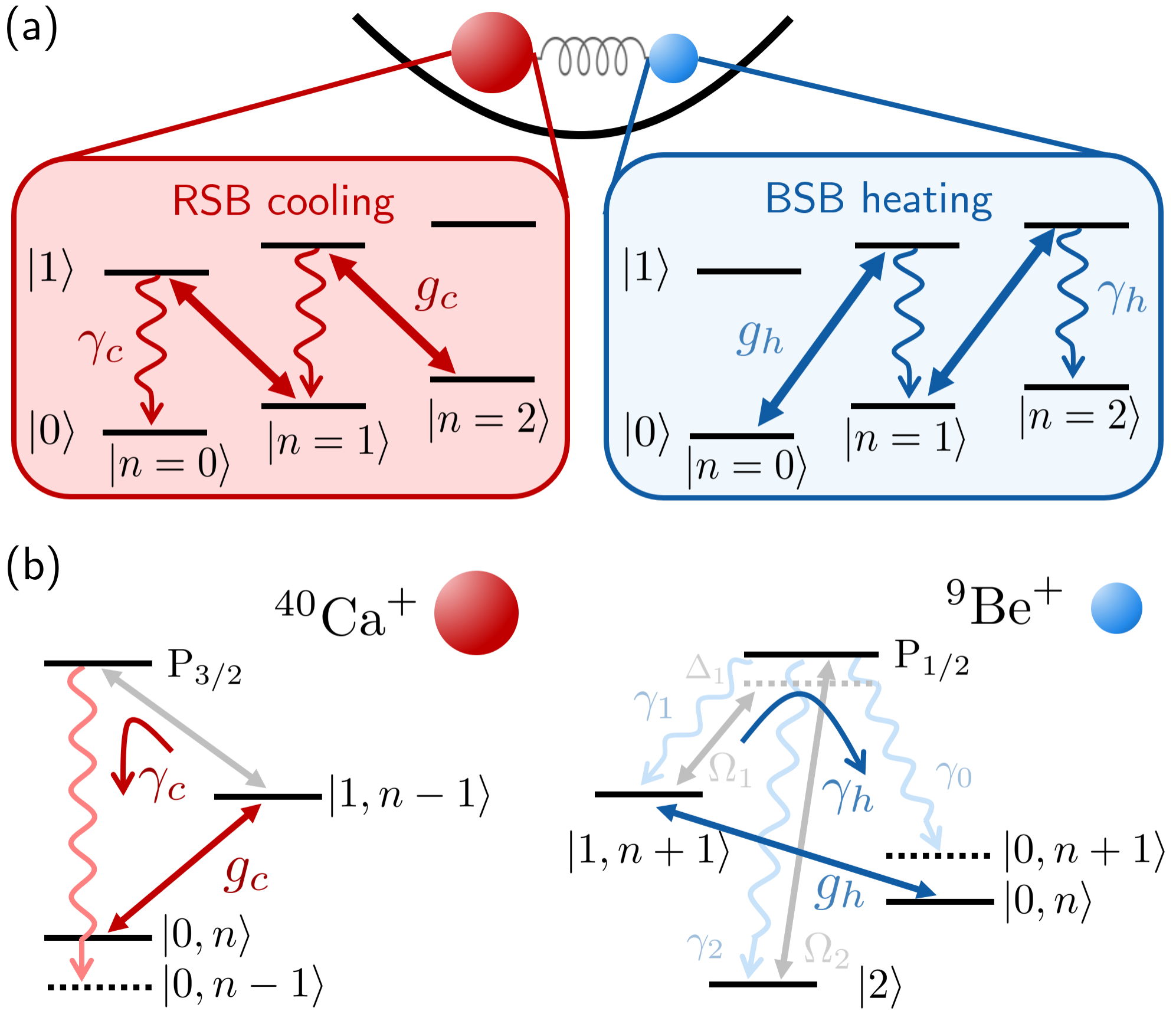}
    \caption{(a) Schematic of the two dissipative channels. The red (blue) sideband plus engineered decay is applied to calcium (beryllium) and thus realizes cooling (heating) of the shared motional mode. (b) Sketch of heating and cooling process including the engineered decay in calcium and beryllium.}
    \label{fig:schematic}
\end{figure}

In this Letter, we implement competing dissipation channels with low cross-talk using a two-species ion chain.  We use this to investigate the realization of a phonon laser which differs from earlier implementations by operating in the resolved-sideband regime at phonon occupations of $\bar{n}<10$.  This system exhibits three distinct phases, one of which corresponds to phonon lasing. We characterize these by extracting the number state distributions \cite{Meekhof1996}. In the lasing phase, we demonstrate phase locking and phase diffusion, diagnosing these effects by measuring the characteristic function of the motional phonon mode \cite{ChristaCharacter}.

\begin{figure}
    \centering
    \includegraphics[width=0.48\textwidth]{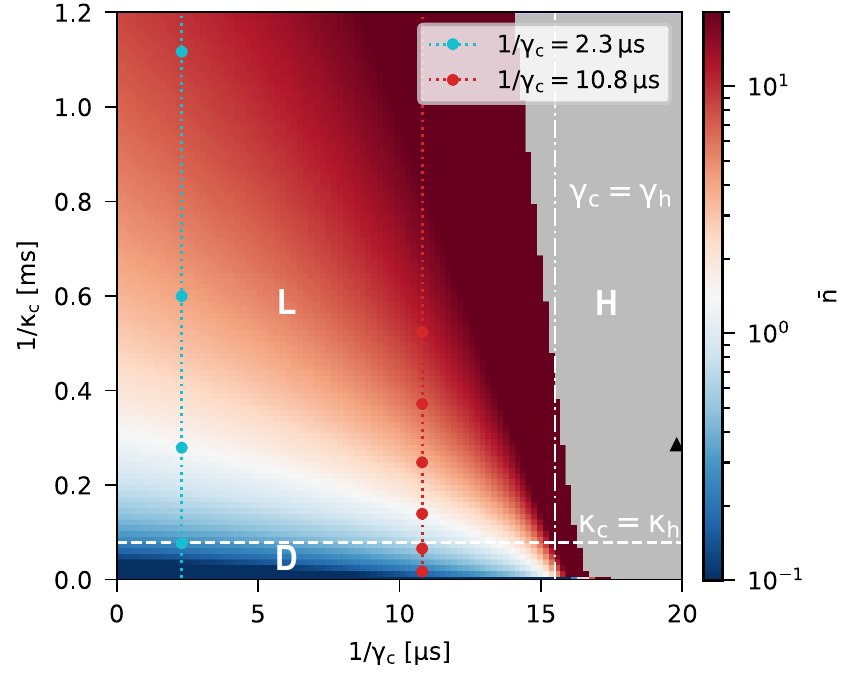}
    \caption{Simulated phase diagram plotted as a function of control parameters, with the average phonon number used to characterize the steady state. The values are generated using numerical simulations including up to 100 motional states, solving for the steady state. $\gamma_\mathrm{c}$ is the internal state damping rate in the cooling ion while $1/\kappa_\mathrm{c}=\gamma_\mathrm{c}/g_\mathrm{c}^2$, with the sideband coupling strength $g_\mathrm{c}$. An effective beryllium decay $1/\gamma_\mathrm{h}=\SI{15.5}{\micro\second}$ and $1/\kappa_\mathrm{h}=\SI{0.08}{\milli\second}$ are fixed. Red and light blue dots correspond to the data taken and shown in Fig.~\ref{fig:data1}. White lines indicate the expected phase transitions at $\gamma_\mathrm{c}=\gamma_\mathrm{h}$ and $\kappa_\mathrm{c}=\kappa_\mathrm{h}$. Values of $\bar{n}>80$ are greyed out, they correspond to heating or high $\bar{n}$  lasing values.}
    \label{fig:phasediagram}
\end{figure}

\begin{figure}
    \centering
    \includegraphics[width=0.48\textwidth]{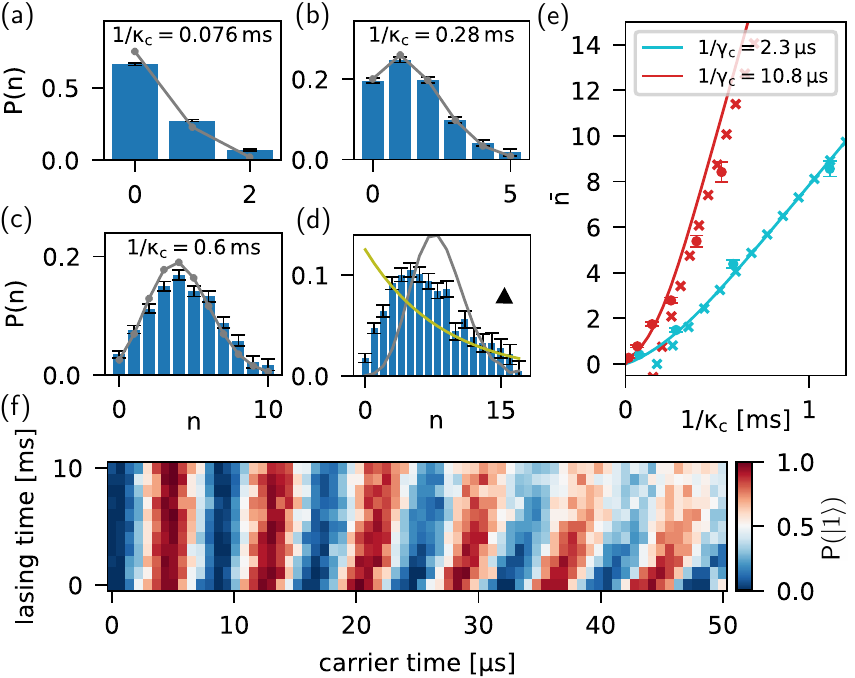}
    \caption{Phonon distributions for different $1/\kappa_\mathrm{c}$. (a)-(c) taken at $1/\gamma_\mathrm{c}=\SI{2.3}{\micro\second}$, show an increase in the average phonon number with an increase of $1/\kappa_\mathrm{c}$. The measured phonon distributions (blue bars) agree well with a Poisson distribution of same $\bar{n}$ (grey). (d) is taken in the heating region at $1/\kappa_\mathrm{c}=\SI{0.28}{\milli\second}$ and $1/\gamma_\mathrm{c}=\SI{19.9}{\micro\second}$ as indicated by the black triangle. Here, a mixture of coherent (grey) and thermal (green) distribution is observed. (e) Average phonon number as function of $1/\kappa_\mathrm{c}$, mean-field theory (crosses), simulation (line) and experiment (dots with error bars) obtained from the phonon distributions. (f) Calcium carrier oscillations taken in the heating region corresponding to (d). The decrease in calcium carrier frequency for increasing lasing times shows the heating nature.}
    \label{fig:data1}
\end{figure}

Our setup \cite{Negnevitsky2018} takes advantage of the spectral separation for the resonant transitions of beryllium relative to calcium ions, which allows excellent individual addressing for both coherent and dissipative optical pumping control fields. Coupling of the internal states to the shared motional mode is provided by a Jaynes-Cummings (JC) Hamiltonian $\hat{H}_\mathrm{c} = \hbar g_\mathrm{c} \left(\create \hat{\sigma}_-^c + \destroy \hat{\sigma}_+^c \right)$ applied to the optical qubit in the calcium ion, and an Anti-Jaynes-Cummings (AJC) Hamiltonian applied to a hyperfine qubit in beryllium $\hat{H}_\mathrm{h} = \hbar g_\mathrm{h} \left(\create \hat{\sigma}_+^h + \destroy \hat{\sigma}_{-}^h \right)$, where $\hat{a}$ is the annihilation operator of the motional mode and $\hat{\sigma}_-^\mathrm{h/c}$ the lowering operator of the internal state of the heating/cooling ion. The dissipation channels described by the Lindblad jump operators $\hat{L}_\mathrm{h/c} = \sqrt{\gamma_\mathrm{h/c}} \hat{\sigma}_-^\mathrm{h/c}$ are implemented using optical pumping via ancillary short-lived states. The combination of coherent and dissipative driving leads to competing motional cooling (index c) from the calcium ion and motional heating (index h) from beryllium as illustrated in Fig.~\ref{fig:schematic} (a). We note that while the gain in our system comes from the use of an AJC Hamiltonian for beryllium allied to internal state pumping from a higher to a lower energy level, a re-labelling of internal state levels produces a JC Hamiltonian plus internal state population inversion, as it is commonly found in lasers.

To understand the system, we compare it with a mean-field model which makes the assumption of a separation of timescales between fast dynamics in the internal states and slower evolution of the oscillator. The resulting equation for $A=\left<\hat{a}\right>$ (for which the derivation is given in the Supplemental Material Sec.~\ref{SM:mean-field}) is
\begin{align}
    \frac{d}{dt}A(t) = A(t)\left(\frac{2\kappa_\mathrm{h}}{1+s_\mathrm{h} \abs{A(t)}^2}-\frac{2\kappa_\mathrm{c}}{1+s_\mathrm{c} \abs{A(t)}^2}\right),
    \label{eq:A}
\end{align}
with the gain/loss coefficients $\kappa_\mathrm{h/c}=g_\mathrm{h/c}^2/ \gamma_\mathrm{h/c}$, and saturation coefficients $s_\mathrm{h/c}= 8\kappa_\mathrm{h/c}/\gamma_\mathrm{h/c}$ which govern the saturation of the atomic inversion $\propto 1/(1+s_\mathrm{h/c} \abs{A(t)}^2)$. This equation allows us to identify separate behaviours which are consistent with a more complete numerical modelling which produces the steady state phase diagram shown in Fig.~\ref{fig:phasediagram}. For $G=\kappa_{\rm h}/\kappa_{\rm c}<1$, the cooling rate of calcium overcomes any heating and the ground state is a steady state. For $s_{\rm h}/s_{\rm c}>G$ (which is equivalent to $\gamma_\mathrm{h}/\gamma_\mathrm{c}<1$) it is the only steady state (``dark'' region (D) in Fig.~\ref{fig:phasediagram}), while for $s_{\rm h}/s_{\rm c}<G$ (equivalent to $\gamma_\mathrm{h}/\gamma_\mathrm{c}>1$) an additional set of parameters produces $\dot{A} = 0$ but this regime is unstable to fluctuations. Where the gain exceeds loss ($G>1$), the system is above threshold and the ground state is unstable. Here two distinct regimes exist. The first, occurring for $s_{\rm h}/s_{\rm c}>G$ is analogous to lasing (L), in which saturation of the heating process leads to a stable steady state for finite excitation $|A|^2$. The second regime occurs when the cooling ion saturates at lower excitations than the heating ion $s_{\rm h}/s_{\rm c}<G$, which leads to a runaway heating effect (H). This latter behaviour is not observed in a standard continuous-wave laser system, where the loss is due to leakage from the laser cavity and does not saturate, but has been postulated for driven-dissipative systems dominated by spin decay \cite{Reiter2020}. 

We examine these regimes experimentally by applying appropriate Hamiltonians and dissipation until the system attains the steady state, and subsequently characterizing the resulting states through reconstruction of the phonon number distribution. Coherent control is implemented in calcium using a narrow-linewidth laser near resonance with the red sideband of the motional common mode $\omega_m = 2\pi\times$\SI{1.8}{\mega\hertz} sideband of the quadrupole transition $\ket{0}\equiv\ket{S_{1/2}, M_J=+1/2}\leftrightarrow\ket{1}\equiv\ket{D_{5/2}, M_J=+3/2}$ at \SI{729}{\nano\meter}. The coupling coefficient in the Lamb-Dicke approximation is given by $g_\mathrm{c} = \eta_\mathrm{c} \Omega_\mathrm{c}$ with the Rabi frequency $\Omega_\mathrm{c}$ proportional to the electric field gradient of the laser light and the Lamb-Dicke parameter $\eta_\mathrm{c} = 0.05$. For beryllium the coupling is produced by a Raman transition near resonance with the blue motional sideband of the qubit transition $\ket{0}\equiv\ket{S_{1/2},F=2, M_F=+2}\leftrightarrow\ket{1}\equiv\ket{S_{1/2},F=1, M_F=+1}$. Here, $g_\mathrm{h} = \eta_\mathrm{h} \Omega_\mathrm{h}$ where $\Omega_\mathrm{h}$ is proportional to the product of the electric fields of the two Raman lasers and the Lamb-Dicke parameter $\eta_\mathrm{h} = 0.15$.

The dissipation channel $\hat{L}_\mathrm{h/c} = \sqrt{\gamma_\mathrm{h/c}} \hat{\sigma}_-^\mathrm{h/c}$ in both ions is provided by laser-driven optical pumping of dipole transitions, see Fig.~\ref{fig:schematic} (b). For calcium, this is performed using a resonant laser field with a wavelength of \SI{854}{\nano\meter} which primarily couples $\ket{1}$ to $\ket{P_{3/2}, M_J = 3/2}$, from where decay to $\ket{0}$ occurs by spontaneous emission. A low fraction of leakage to other levels is mitigated by additional repumping laser fields at \SI{397}{\nano\meter} and \SI{866}{\nano\meter}, not shown.  For beryllium, optical pumping from $\ket{1}$ to $\ket{0}$ uses circularly polarized light which couples $\ket{1}$ to the short-lived $P_{1/2}$ excited state. The branching ratio for decay from the excited $P$ state into $\ket{0}$ is only one third, with additional decay back into $\ket{1}$ and $\ket{2}=\ket{S_{1/2}, F = 2, M_F = +1}$. We use an additional strong \SI{313}{\nano\meter} beam resonant with $\ket{2} \rightarrow \ket{P_{1/2}, F=2, M_F=2}$ to repump this population and detune the $\ket{1} \rightarrow P_{1/2}$ laser by \SI{10}{\mega\hertz} to avoid the formation of a coherent dark state. The 4-level nature of the repumping in beryllium results in an additional dephasing of the $\ket{0}$ and $\ket{1}$ states which is not captured by $\hat{L}$ in the two-level description. We find that the effect of the four levels is captured by an additional dissipation operator $\hat{L}_1 = \sqrt{\gamma_\mathrm{e}}\ket{1}\bra{1}$ at rate $\gamma_\mathrm{e}=f\gamma_\mathrm{h}$ (with $f=50/40$ representing the ratio of decay strengths from the $P$ state to $\ket{1}$ rather than to $\ket{0}$). In the equations above this has the effect of modifying $\kappa_\mathrm{h}$ to $\kappa_\mathrm{h} = g_\mathrm{h}^2/(\gamma_\mathrm{e}+\gamma_\mathrm{h})$. All numerical simulations displayed use the full 4-level beryllium dynamics.

\begin{figure}
    \centering
    \includegraphics[width=\columnwidth]{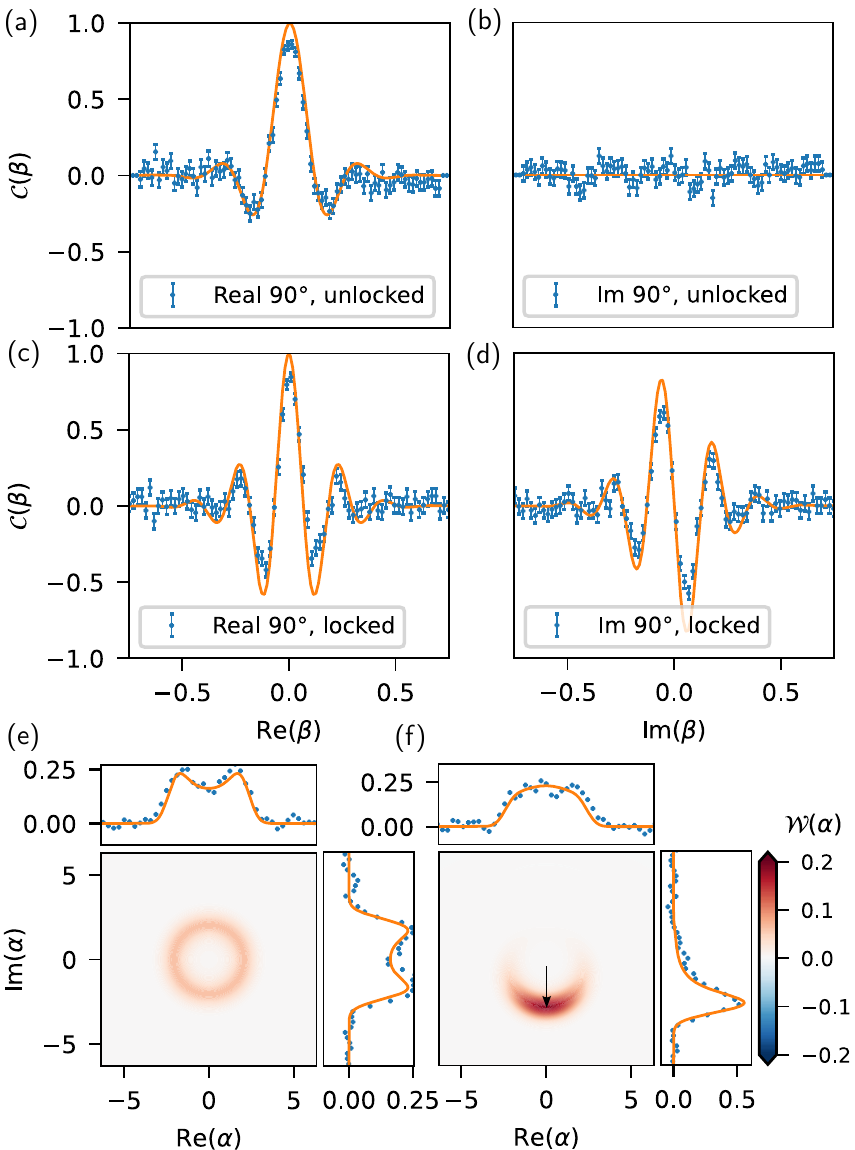}
    \caption{(a)-(d) Simulation (orange line) and experimental data (blue dots) of real and imaginary part of the characteristic function along the 90 degree axis in the unlocked (a, b) and locked (c, d) case. Symmetry breaking in the imaginary part (b) to (d) is observed. (e)-(f) Simulated Wigner function (2D) and marginal probability (side plots, orange line) along two axes for the (e) unlocked and (f) phase locked phonon laser including experimental data (blue dots).}
    \label{fig:phase_lock}
\end{figure}

In experiments, we fix the parameters for the beryllium (heating) ion, choosing to vary $\kappa_{\rm c}, s_{\rm c}$ through control of calcium. State reconstruction is performed using standard methods, utilizing the number-state dependence of Rabi oscillations on the motional sidebands of the calcium ion \cite{Meekhof1996, Kienzler2015}. We take two slices of the phase diagram using settings which take two distinct values of $1/\gamma_\mathrm{c}$. For each value, we extract phonon distributions for several values of $1/\kappa_\mathrm{c}$ (y-axis of phase diagram). Observed phonon distributions are presented in Fig.~\ref{fig:data1} (a)-(d), with the obtained mean phonon number shown in Fig.~\ref{fig:data1} (e). We note that the latter increases as $\kappa_{\rm h}/\kappa_{\rm c}$ transitions from being below to above one, indicating a change from a dark to lasing steady state. Agreement is observed with respect to numerical simulations (continuous line) using the calibrated parameters and with respect to predictions of mean-field calculations (crosses) -- these parameters and the calibration methods are given in the Supplemental Material Sec.~\ref{SM:exp_params} and Sec.~\ref{SM:engineered_decay}.  The obtained phonon distribution in (a) exhibits a peak in $P(n)$ at the origin, whereas in (b) and (c) the distributions are close to those of a Poisson distribution (shown in grey) with the same mean phonon number, which is expected from a coherent state generated by lasing. We also observe the increased saturation of the heating ion (beryllium) as the system transitions to lasing, for instance in the $1/\gamma_\mathrm{c}=\SI{2.3}{\micro\second}$ data set the measured spin population of the steady state increases from $\langle \hat{\sigma}_z^h\rangle=-0.60$ to $\langle \hat{\sigma}_z^h\rangle=-0.26$ as $\kappa_c$ is decreased.

As $\gamma_\mathrm{c}$ is decreased relative to $\gamma_\mathrm{h}$, we  reach the heating phase in which the phonon distribution does not attain a steady state. Data from a phonon distribution taken in this regime and sampled at a finite time is shown in Fig.~\ref{fig:data1} (d) and shows a mixture of a coherent (grey) and thermal (green) distribution. Since the system does not reach a steady state, we do not only reconstruct the phonon distribution after a finite time but accompany this by probing Rabi oscillations on the $\ket{0}\leftrightarrow\ket{1}$ carrier transition (for which the Rabi rate is dependent on the phonon occupation \cite{bible}) as a function of the dissipation time - data of this type is shown in Fig.~\ref{fig:data1} (f). Here we observe continued slowing of the Rabi oscillation and reduction of coherence as a function of time, indicating that the mean and variance of the phonon distribution increases.

One experimental issue which impeded initial attempts to understand the dissipative dynamics was heating of spectator motional modes, which we suspect is caused by photon scattering during recoil in internal state repumping.
This in turn modifies the laser-ion interaction on the motional mode used for the laser. To mitigate this, we interleave the lasing pulse with short periods of sideband cooling on the spectator out-of-phase axial mode performed using the beryllium ion. The repetition frequency was chosen to be low enough that the internal states of the ion could re-attain a steady state during the dual-dissipation drive.

\begin{figure}
    \begin{tabular}{c}
    \includegraphics[width=0.98\columnwidth]{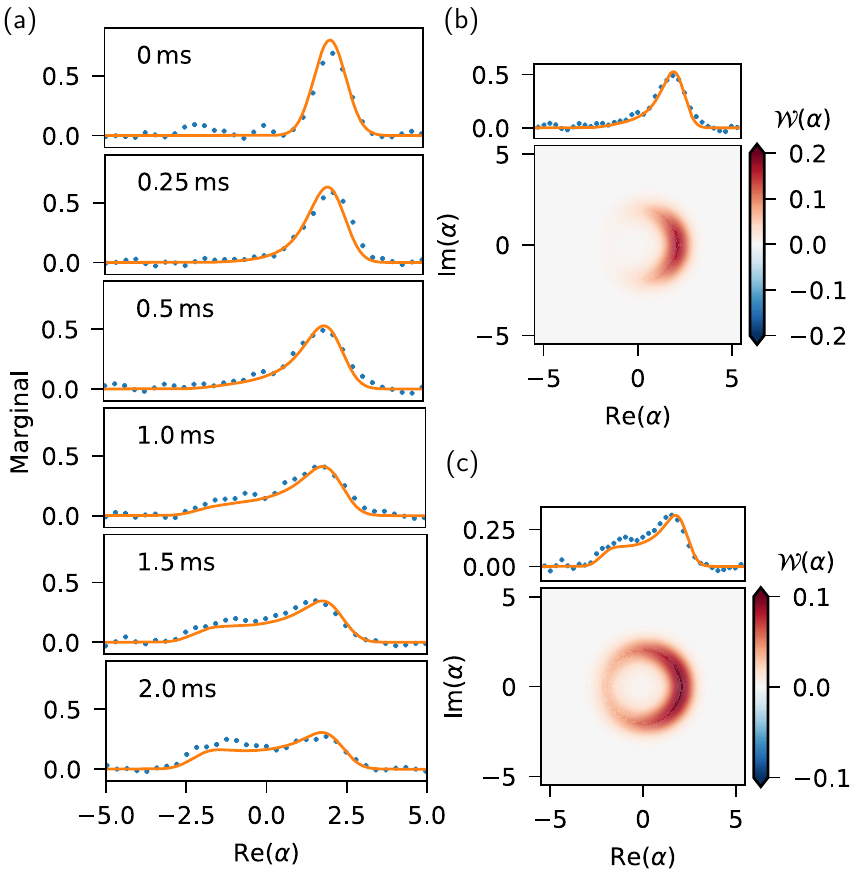}
    \end{tabular}
    \caption{Phase diffusion. (a) Marginal probabilities along 90 degree axis for different lasing lengths. Starting from a coherent state with a well-defined phase, the lasing process introduces phase diffusion over time (simulation in orange, experiment in blue). (b) and (c) Simulated Wigner functions (2D) with marginal probabilities for a dissipation duration of 0.5ms and 1.5ms.}
    \label{fig:phase_diffusion}
\end{figure}

In addition to probing the number state distributions, we also measured the characteristic function $\mathcal{C}(\beta)$, which fully characterizes the oscillator state. We measure this by coupling the oscillator to the internal state of the ion using a  state-dependent force followed by a projective spin readout \cite{ChristaCharacter}. This measurement is performed after turning off the lasing dynamics and repumping the internal state of the calcium ion to $\ket{0}$.  The measured characteristic function for the lasing state is shown in Fig.~\ref{fig:phase_lock} (a) and (b). Fig.~\ref{fig:phase_lock} (e) shows the marginals of the Wigner function $\mathcal{W}(\alpha)$ for a state with $\bar n \approx 4.4$, obtained from the characteristic function by a discrete Fourier transform. These are accompanied by the results of a steady state numerical solution of the Lindblad master equation. The characteristic function and Wigner function are symmetric about zero, which indicates the random phase of the laser. The phase symmetry can be broken by adding an external drive of a well-defined phase to which the oscillator locks. We introduce such a drive by applying a resonant oscillating force to the oscillator using a voltage applied to a trap electrode, resulting in a Hamiltonian 
$\hat{H}_t = \hbar g (\hat{a}e^{i\Phi_t}+\hat{a}^\dagger e^{-i\Phi_t})$. 
The force strength $g=2 \pi\times \SI{0.4}{\kilo\hertz}$ is chosen to be weak enough such that it does not overwhelm the lasing dynamics. Fig.~\ref{fig:phase_lock} (c) and (d) show data for the characteristic function measured with the force on. Fig.~\ref{fig:phase_lock} (f) shows the corresponding marginals of the Wigner function. The experimental data agree well with numerical simulations performed using experimentally observed parameters, which are plotted alongside. Clear symmetry breaking in the imaginary part of the characteristic function is observed from Fig.~\ref{fig:phase_lock} (b) to (d). From comparison with the simulated Wigner function, we see that the state is nevertheless not perfectly represented by a coherent state, with some phase diffusion being visible. 

To investigate phase diffusion during the dissipation, we examine the effects on an initial coherent state which has the same $\bar{n} = 4.4$ as the steady state but a well-defined phase. We prepare this by ground-state cooling followed by the application of an oscillating force for a pre-calibrated duration. After applying the dissipation drives for a fixed duration, we reconstruct the characteristic function. Fig.~\ref{fig:phase_diffusion} (a) shows the marginals of the Wigner distribution along the real axis, which show the effects of diffusion. Results from numerical simulations (orange line) using a 4-level beryllium system and 2-levels for calcium with parameters obtained using independent calibration experiments agree with the data (blue dots), and allow us to extract a linear increase in the phase uncertainty with time at a rate $\dot{\langle \theta^2 \rangle} = 0.022~{\rm rads}^2/s$. This diffusion is analogous to that which produces the Schawlow-Townes limit for the laser linewidth \cite{ShawlowTownes, scully_zubairy}. 

To obtain simple models of the physics, it is helpful to reduce the problem to a pair of two-level systems interacting with a single mode. We explored this approximation initially in numerical models using Lindblad operators $\hat{L}_\mathrm{h}$, $\hat{L}_\mathrm{c}$.
However, without the presence of the additional dephasing operator $\hat{L}_1 = \sqrt{\gamma_\mathrm{e}}\ket{1}\bra{1}$ (accounting for additional dissipation processes in the 4-level beryllium ion) we found that the mean mode occupation and diffusion were significantly overestimated (for the experimental parameters of the data in Fig.~\ref{fig:phase_diffusion} it produces $\bar n = 5.3$ and $\dot{\langle \theta^2 \rangle} = 0.054~{\rm rads}^2/s$). Introduction of the dephasing operator $\hat{L}_1$ produced much better agreement of the mean phonon number and also resulted in reasonable predictions for the saturation of beryllium and the phase diffusion in the lasing phase. Analytical results based on Eq.~\eqref{eq:A} reproduce the mean phonon number and saturation values. For the phase diffusion, we applied the  Heisenberg-Langevin theory \cite{scully_zubairy} to our problem including $\hat{L}_1$ (details of the calculation can be found in the Supplemental Material Sec.~\ref{SM:phasediff_HL} and \ref{SM:phasediff_HL_wdephasing}), but find that this seems to still overestimate the diffusion coefficient by a significant factor compared to the numerical simulations (in the case above, it produces a value of $0.041~{\rm rads}^2/s$).
We do not have an intuition why such a discrepancy might exist. All models and numerical simulations use finite Lamb-Dicke parameters for beryllium, while we used a first-order expansion approximation for calcium -- we verified that this approximation had a minor effect on the theoretical predictions for our parameter regimes.  

The work performed here differs in two important aspects from earlier work on phonon lasing with trapped ions \cite{Vahala2009}. First, the phonon occupation is lower, stemming from the use of the resolved-sideband regime. Hence, our results sit at the boundary where quantum effects are accessible and methods are available for extracting this information, such as characteristic function and Fock-state reconstruction. Second, there are fundamental differences in the physics. In earlier work, which used a transition in a single ion driven simultaneously at two frequencies, the balance between heating and cooling appeared due to non-linearity in the ion-light interaction caused by the modification of the lineshape for large Doppler shifts, with no saturation of the internal state. Here, by contrast, the saturation of the ``gain'' ion is what introduces non-linearity. 

The experiments above use a relatively simple pair of dissipation channels, involving single-frequency laser tones applied in the Hamiltonian part of the control. 
By using multi-frequency drives, similar experiments could explore reservoirs for which the underlying state space is more easily described in a squeezed or displaced Fock basis.
This extension would provide a realization of a squeezed phonon laser, which has been predicted to offer advantages in sensing \cite{Porras, Ivanov_2020, Wei2022}. Extensions of competing reservoir systems to multiple ions would allow for the realization of more generalized spin-boson models such as the open Dicke model.

TB and TLN performed the experiments, with experimental support from BN, MS, MM and FL. Theoretical preparatory work was  performed by TLN and FR with support from JH. Data analysis was performed by TB and TLN. Theoretical understanding of the results was produced by DB, TB, FR, SFY, JH. The paper was written by TB and JH with input from all authors.

We acknowledge support from the Swiss National Science Foundation (SNF) under grant number 200020 179147/1, the 
SNF through the National Centre of Competence in Research for Quantum Science and Technology (QSIT) grant 51NF40–185902, the SNF Ambizione grant PZ00P2-186040 and the EU Quantum Flagship H2020-FETFLAG-2018-03 grant 820495 AQTION. SFY would like to acknowledge funding from the NSF through the CUA PFC.


%% file: supplemental_material.tex
\title{Supplemental material for: \\ \thetitle}
\setcounter{footnote}{34}
\theauthors
\maketitle

\onecolumngrid
\renewcommand{\theequation}{S.\arabic{equation}}
\renewcommand{\thefigure}{S.\arabic{figure}}
\setcounter{table}{0}
\renewcommand{\thetable}{S.\arabic{table}}

\section{Experimental Parameters}
\label{SM:exp_params}
A variety of experimental parameters were used across the data and theory presented in the main text. These are given in Table.~\ref{parameters}. 
For inputs to simulations we chose values close to those accessed by the experiments. 
For values obtained from experiment, we quote values measured in calibration experiments including statistical error bars. These include the use of Rabi oscillations to extract Rabi oscillation rates and optical pumping experiments with exponential decay features used to extract damping parameters. 

Calibration of the lasing experiments include the calibration of the red and blue sideband frequencies, the three decay parameters as well as the Stark shift calibrations for the simultaneous laser pulses during the experiment. The calibration process takes up to 30 minutes.
Taking data for phonon distributions with high average phonon numbers (sampling larger frequencies) requires averaging of relatively many data points leading to experiments lasting up to 25 minutes. Therefore for one scan, calibration and data taking together takes up to one hour. Sequential calibrations taken one hour apart (i.e. at the beginning of a calibration procedure and after the scan was finished) showed drifts of up to about 1.5 times the statistical error bar.
Phase locking experiments using the SDF readout are faster but require a longer calibration procedure, including a precise trap frequency calibration as well as a tickling phase calibration, resulting in similar timescales for calibrations and data taking, hence they exhibit similar drifts to the other experiments. 


\begin{table}[h]
    \begin{center}
    \begin{tabular*}{0.8\linewidth}{@{\extracolsep{\fill}}l c c c c c}
    \hline
    \hline
    & $g_\mathrm{h}/2\pi$ [\SI{}{\kilo\hertz}]
    & $g_\mathrm{c}/2\pi$ [\SI{}{\kilo\hertz}] &
    $\gamma_{h1}$ [\SI{}{\per\milli\second}] & $\gamma_{h2}$ [\SI{}{\per\milli\second}] & $\gamma_\mathrm{c}$ [\SI{}{\per\milli\second}]\\
    \hline
    Fig. 2  & $4.55$  & - & 91 & 435 & -  \\
    Fig. 3a  & $4.65\pm0.02$  & $12.0\pm0.1$ & $96\pm6$ & $435\pm38$  & $426\pm11$  \\
    Fig. 3b  & $4.62\pm0.03$  & $6.28\pm0.05$ & $91\pm7$ & $385\pm44$  & $429\pm13$  \\
    Fig. 3c  & $4.65\pm0.02$  & $4.29\pm0.04$ & $96\pm6$ & $435\pm38$  & $426\pm11$  \\
    Fig. 3d  & $4.57\pm0.02$  & $2.11\pm0.01$ & $93\pm6$ & $385\pm30$  & $50\pm2$   \\
    Fig. 3e  & $4.63$  & - & 93 & 435  & -   \\
    Fig. 4  & $4.59$  & $4.24$ & 91 & 344  & 435  \\
    Fig. 5  & $4.59$  & $4.24$ & 91 & 344  & 435  \\
    \hline
    \hline
    \end{tabular*}
    \caption{\label{parameters} Primary experimental parameters in the presented experimental data and simulations. The parameter $\gamma_{h1/h2}$ corresponds to $1/\tau_{1/2}$ as defined in Sec.~\ref{SM:engineered_decay}.}
    \end{center}
\end{table}

\section{Engineered Decay}
\label{SM:engineered_decay}
As described in the main text and illustrated in Fig.~\ref{fig:schematic}, optical pumping in calcium is performed from $\ket{1}$ to $\ket{0}$ using a resonant laser field at \SI{854}{\nano\meter} which primarily excites to $\ket{P_{3/2}, M_J = 3/2}$, from where decay to $\ket{0}$ occurs by spontaneous emission. Leakage to other levels is mitigated by additional repumping laser fields at \SI{397}{\nano\meter} and \SI{866}{\nano\meter}. We tune the effective repumping strength from $\ket{1}$ to $\ket{0}$ by adjusting the power of the \SI{854}{\nano\meter} light. For calibration, we prepare the $\ket{1}$ state and measure population of $\ket{0}$ as a function of time with the \SI{854}{\nano\meter} and additional fields turned on. The effective decay rate can then be determined from an exponential fit which we use to model calcium as an effective 2-level system after adiabatically eliminating the P-state.

For beryllium, the 4-level system consisting of $\ket{0}$, $\ket{1}$, $\ket{2}=\ket{S_{1/2}, F = 2, M_F = +1}$ and the P-state $\ket{e}=\ket{P_{1/2}, F=2, M_F=2}$ requires a more complex treatment. This involves the two repump lasers: repumper 1 from $\ket{1 }$ to $\ket{e}$ with Rabi frequency $\Omega_1$, repumper 2 from $\ket{2}$ to $\ket{e}$ at Rabi frequency $\Omega_2$, and subsequent spontaneous emission from $\ket{e}$ into all three levels $\ket{0}$, $\ket{1}$ and $\ket{2}$ at rates $\gamma_0, \gamma_1, \gamma_2$ respectively. This arrangement realizes an effective decay from $\ket{1}$ to $\ket{0}$ with rate $\gamma_\mathrm{h}$. Repumper 1 is red detuned by $\Delta_1 = 2 \pi \times 10$MHz to avoid formation of a coherent dark state. While the effective decay rate from $\ket{1}$ to $\ket{0}$ with both repumpers turned on could be measured analogously to calcium, we are interested in determining the two repumping strengths $\Omega_1$, $\Omega_2$, individually in order to model our system with all four levels. In the following, we describe how we obtain $\Omega_1$ and $\Omega_2$.

Analytically, we can describe the beryllium 4-level system with with only repumper 1 turned on with the following rate equations:
\begin{align}
	\dot{P_0}&=\gamma_0 P_\mathrm{e}\\
    \dot{P_1}&=-b P_1+\gamma_1 P_\mathrm{e}\label{rho0}\\
	\dot{P_2}&=\gamma_2 P_\mathrm{e}\\
	\dot{P_\mathrm{e}}&=b P_1-\gamma P_\mathrm{e}\overset{!}{=}0\label{rhoe}.
\end{align}
where the $P_j$ denote the populations of levels $\ket{j}$, $b$ is the rate at which the repumper 1 laser excites from $\ket{1}$ to $\ket{e}$ and $\gamma \equiv \gamma_0+\gamma_1+\gamma_2$. Assuming that we can adiabatically eliminate the excited state, from Eq.~\eqref{rhoe} we obtain
\begin{align}
	P_\mathrm{e}&=b P_1/\gamma.
\end{align}
Inserting this into the other equations, and solving, we obtain
\begin{align}
	P_1(t) &=e^{-b\frac{\gamma-\gamma_1}{\gamma}t}=e^{\frac{-b(\gamma_0+\gamma_2)}{\gamma}t}\\
	P_0(t) &=\frac{-\gamma_0}{\gamma_0+\gamma_2}e^{\frac{-b(\gamma_0+\gamma_2)}{\gamma}t}=\frac{-\gamma_0}{\gamma_0+\gamma_2}e^{-\frac{1}{\tau_1}t}\label{tau1}
\end{align}
where $1/\tau_1=b(\frac{\gamma_0+\gamma_2}{\gamma})$ is the the inverse of the time constant describing the effective repumping to the $\ket{0}$ state.

To find an expression for $b$, we compare these rate equations to results obtained by applying the effective operator formalism \cite{ReiterEffective} to the Master equation for the 4-level system. Eliminating the short-lived P-level state $\ket{e}$, we obtain a set of effective Lindblad operators $L_\text{eff}$ and identify the matrix entry which produces the appropriate pumping from $\ket{1}$ back to $\ket{1}$. The square of this entry then describes the rate of the pumping process. By setting this matrix element equal to $b\gamma_1/\gamma$ we can obtain $b$.

Considering the case where only repumper 1 is turned on, this leads to 
\begin{equation}
	b=\gamma\left(\frac{\Omega_1^2}{\gamma^2+4\Delta^2}\right).
\end{equation}
We can use this in the expression of $\tau_1$ and solve for $\Omega_1$,
\begin{equation}
    \Omega_1= \sqrt{\frac{1}{\tau_1}\frac{\gamma^2+4\Delta^2}{\gamma_0+\gamma_2}}.
\end{equation}

We obtain $\tau_1$ from calibration experiments in which we start from the $\ket{1}$ state and turn on repumper 1, measuring the rate at which the $\ket{0}$ state is populated. An additional shelving pulse is applied to $\ket{2}$ before detection to prevent population from this state producing a mis-diagnosis of $\ket{0}$ population due to off-resonant scattering. This experiment also allows us to observe saturation of the bright state population at $57\%$, which is consistent with the expected branching ratio.

Analogously, starting from rate equations where only repumper 2 is assumed to be turned on we obtain an expression for the strength of repumper 2:
\begin{equation}
	\Omega_2= \sqrt{\frac{1}{\tau_2}\frac{\gamma^2}{\gamma_0+\gamma_1}}.
\end{equation}

We obtain $\tau_2$ by measuring the rate at which the $\ket{0}$ state is populated after initialization in $\ket{2}$ and with repumper 2 turned on. In this case the bright state saturates at $43\%$. The preparation of state $\ket{2}$ is performed by optically pumping into $\ket{0}$ and then pumping into $\ket{2}$ by 10 repetitions of a cycle involving a coherent transfer pulse to $\ket{1}$ plus application of the repump laser 1.

With our model including the four levels, we also calculate that the repumping from $\ket{1}$ to $\ket{0}$ can be approximated by an effective exponential with rate $\gamma_{\mathrm{h}}=1/15.5~\mu\text{s}^{-1}$. For the P-state we calculate the decay rates on different paths using the Clebsch-Gordon coefficients and the measured decay rate of the upper state \cite{BePeLifetime}, and find them to be $\gamma_{0}=40.0~\mu\text{s}^{-1}$, $\gamma_{1}=50.4~\mu\text{s}^{-1}$ and $\gamma_{2}=29.6~\mu\text{s}^{-1}$.

\section{Experimental Methods: Trap Frequency, SDF and Tickling Strength}
The experimental methods and calibrations of the trap frequency, SDF, and tickle strengths were the same as described in \cite{deNeeve2022}.

\section{Experimental Methods: Characteristic Function and Marginals}
In order to probe the phase of our phonon laser, we read out the real and imaginary part of the characteristic function \cite{ChristaCharacter} along the two phase space axis using state-dependent forces (SDF). A complex Fourier transformation of the characteristic function then yields the marginals, which are the probability density functions along the position or momentum axis. For the data presented in Fig.~\ref{fig:phase_lock} and Fig.~\ref{fig:phase_diffusion}, the characteristic function was measured from $-0.7$ to $0.7$. Before taking the complex Fourier transformation, we pad the data with zeros from $-1.0$ to $-0.7$ and from $0.7$ to $1.0$.

\section{Mean-field Approach}
\label{SM:mean-field}
In the following, we derive an expression for the phonon number in the mean-field approximation as well as an evolution equation for the phonon mode. \newline
The internal ion transitions $\ket{0}_j \leftrightarrow \ket{1}_j$ ($j \in \{\mathrm{c}, \mathrm{h}\}$) are coupled to the ion motion $a$ by the blue (heating) and red (cooling) sideband drives:
\begin{align}
H_\mathrm{h} &= g_\mathrm{h} ( a^\dagger \sigma^\mathrm{h}_+ + a \sigma^\mathrm{h}_- ),\\
H_\mathrm{c} &= g_\mathrm{c} ( a^\dagger \sigma^\mathrm{c}_- + a \sigma^\mathrm{c}_+ ).
\label{eq:H:c}
\end{align}
Here, $\sigma^\mathrm{c/h}_{+/-}$ denotes the excitation/de-excitation of the internal state of the cooling/heating ion and $a^\dagger$ ($a$) the creation (annihilation) of an excitation of the motional mode. Individual addressing of the two species due to spectral isolation allows for independent coupling constants, $g_\mathrm{c}$ and $g_\mathrm{h}$.\newline
Beside the Hamiltonian couplings, the system is subjected to dissipation by spontaneous emission of the heating and cooling ion, modeled by the jump operators:
\begin{align}
L_\mathrm{h} &= \sqrt{\gamma_\mathrm{h}} \sigma_-^\mathrm{h},
\label{eq:L:h}
\\
L_\mathrm{c} &= \sqrt{\gamma_\mathrm{c}} \sigma_-^\mathrm{c}.
\label{eq:L:c}
\end{align}

The time evolution of the system, which is described by the density operator $\rho$, is governed by a master equation in Lindblad form
\begin{align}
\frac{d \rho}{dt} = \mathcal{L}(\rho) = -i [H, \rho] + \sum_k \mathcal{D}[L_k](\rho).
\label{eq:master}
\end{align}
The Liouvillian $\mathcal{L}(\rho)$ contains a Hamiltonian part with $H = H_\mathrm{c} + H_\mathrm{h}$, and a dissipator
\begin{align}
\mathcal{D}[L_k](\rho) = L_k \rho L_k^\dagger - \frac{1}{2}(L_k^\dagger L_k \rho + \rho L_k^\dagger L_k),
\end{align}
for each jump operator $L_k\in\{L_\mathrm{h},L_\mathrm{c}\}$.\newline
In the Heisenberg picture, the time evolution of the expectation value of a time-independent operator $O$ is described by
\begin{align}
\frac{d}{dt} \expval{O} = \frac{d}{dt} \mathrm{Tr}(\rho O) = \mathrm{Tr}\left(\frac{d \rho}{dt} O\right).
\end{align}
With Eq.~\eqref{eq:master} this can be written as
\begin{align}
\frac{d}{dt} \expval{O} = i \expval{[H,O]} + \sum_k \expval{\tilde{\mathcal{D}}[L_k](O)},
\end{align}
where the dissipative part is given by
\begin{align}
\expval{\tilde{\mathcal{D}}[L_k](O)} = \expval{L_k^\dagger O L_k - \frac{1}{2}(L_k^\dagger L_k O + O L_k^\dagger L_k)}.
\end{align}

To obtain a simple description of the dynamics, we derive the dynamical equations using a cumulant approximation to first order. We choose the variables $A=\expval{a}$ for the ion motion and $S_{c/h}=\expval{\sigma^\mathrm{c/h}_\mathrm{+}}$, $D_\mathrm{c/h}=\expval{\sigma^\mathrm{c/h}_\mathrm{z}}$ for the ion spin:

\begin{align}
    \frac{d}{dt}A(t) &= -ig_\mathrm{c}S^{*}_\mathrm{c}(t) -ig_\mathrm{h}S_\mathrm{h}(t),\label{eq:breuermf:A}\\
    \frac{d}{dt}S_\mathrm{c}(t) &= -\frac{\gamma_\mathrm{c}}{2}S_\mathrm{c}(t)-ig_\mathrm{c}A^{*}(t)D_\mathrm{c}(t),\\
    \frac{d}{dt}D_\mathrm{c}(t) &= 2ig_\mathrm{c}\left(A^*(t)S^*_\mathrm{c}(t) - A(t)S_\mathrm{c}(t)\right) -\gamma_\mathrm{c}\left(D_\mathrm{c}(t)+1\right),\\
    \frac{d}{dt}S_\mathrm{h}(t) &= -\frac{\gamma_\mathrm{h}}{2}S_\mathrm{h}(t)-ig_\mathrm{h}A(t)D_\mathrm{h}(t),\label{eq:mf:dgl:s}\\
    \frac{d}{dt}D_\mathrm{h}(t) &= 2ig_\mathrm{h}\left(A(t)S^*_\mathrm{h}(t) - A^*(t)S_\mathrm{h}(t)\right)-\gamma_\mathrm{h}\left(D_\mathrm{h}(t)+1\right).
    \label{eq:breuermf:Dc}
\end{align}
In order to find an analytic expression for the phonon expectation we make two further assumptions: $\gamma_\mathrm{c} \gg g_\mathrm{c}$ and $\gamma_\mathrm{h} \gg g_\mathrm{h}$. Physically this means that the internal dynamics of the ion reaches a steady state on a time scale which is short compared to the evolution timescale of the ion motion. We can therefore assume the steady state for the spin variables  ($\frac{d}{dt}S_\mathrm{c/h}(t)=0$ and $\frac{d}{dt}D_\mathrm{c/h}(t)=0$) and use the resulting four equations to obtain a single evolution equation for the ion motion,
\begin{align}
    \frac{d}{dt}A(t) = A(t)\left(\frac{\frac{2g_\mathrm{h}^2}{\gamma_\mathrm{h}}}{1+\frac{8 g_\mathrm{h}^2}{\gamma_\mathrm{h}^2} \abs{A(t)}^2}-\frac{\frac{2g_\mathrm{c}^2}{\gamma_\mathrm{c}}}{1+\frac{8 g_\mathrm{c}^2}{\gamma_\mathrm{c}^2} \abs{A(t)}^2 }\right)=A(t)\left(\frac{2\kappa_\mathrm{h}}{1+s_\mathrm{h} \abs{A(t)}^2}-\frac{2\kappa_\mathrm{c}}{1+s_\mathrm{c} \abs{A(t)}^2}\right),
    \label{eq:A:master}
\end{align}
where we have introduced the following notation in the second equation for the gain and loss coefficients: $\kappa_\mathrm{h/c}=\frac{g_\mathrm{h/c}^2}{\gamma_\mathrm{h/c}}$ as well as for the saturation coefficients: $s_\mathrm{h/c}=\frac{8 g_\mathrm{h/c}^2}{\gamma_\mathrm{h/c}^2}$. This concludes the derivation of Eq.~\eqref{eq:A} given in the main text.\newline

\begin{figure}
    \centering
    \includegraphics[width=10cm]{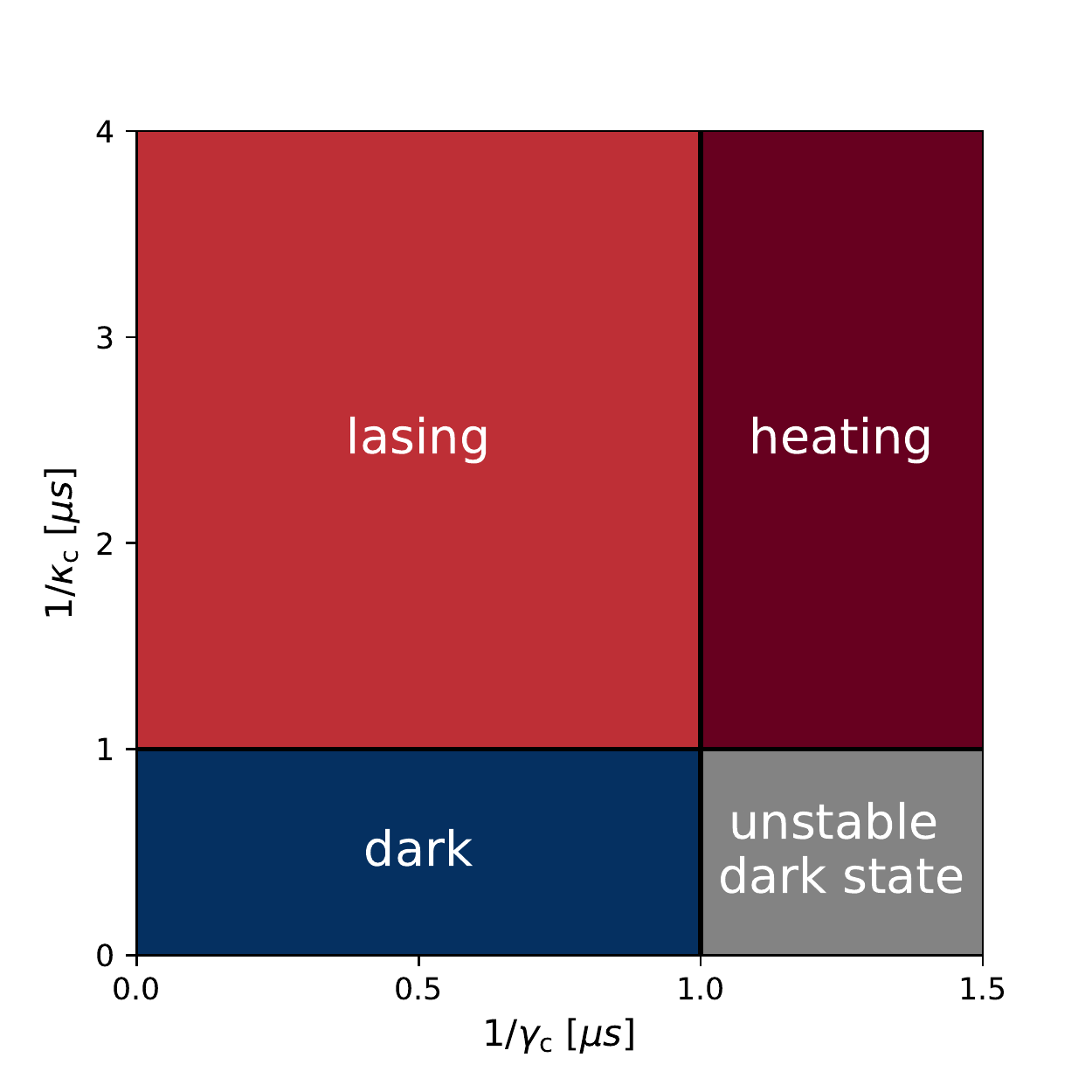}
    \caption{Phase diagram of the phonon laser for fixed $g_\mathrm{h}=\SI{1.0}{\micro\second^{-1}}$ and $\gamma_\mathrm{h}=\SI{1.0}{\micro\second^{-1}}$: the phase space is divided into four phases according to the mean-field theory results. In the lower left corner where $\gamma_\mathrm{h}<\gamma_\mathrm{c}$ and $\kappa_\mathrm{h}<\kappa_\mathrm{c}$ the system is in the dark state. Upon decreasing $\kappa_\mathrm{c}$ with respect to $\kappa_\mathrm{h}$ (which is fixed to $\SI{1.0}{\micro\second^{-1}}$) we cross the lasing threshold and the system assumes a finite coherent phonon occupation (as shown in the experiment) associated with the lasing state. By crossing the vertical line corresponding to $\gamma_\mathrm{h}=\gamma_\mathrm{c}$ we enter the runaway phase where the finite steady states disappear and the system heats up. In the lower right corner we would expect the system to be in the dark state. However, due to the proximity of unstable steady states to this dark state and the presence of fluctuations around this dark state, the system experiences the same runaway behaviour as in the heating phase.}
    \label{fig:phaseDiagram:mf}
\end{figure}

We now discuss the phase transitions and the mean phonon expectation value, which can be extracted from the above equation by analyzing the steady states for $A(t)$.\newline
$A(t)=0$ is clearly a steady state of the system as the time derivative of $A$ vanishes in this case. This state is only stable if the term in the brackets is negative, which corresponds to the condition $\kappa_\mathrm{h}=\frac{g_\mathrm{h}^2}{\gamma_\mathrm{h}}<\frac{g_\mathrm{c}^2}{\gamma_\mathrm{c}}=\kappa_\mathrm{c}$. The condition $\kappa_\mathrm{h}=\kappa_\mathrm{c}$ gives thus the lasing phase transition: upon crossing this transition the zero steady state becomes unstable and $A$ reaches a finite value. This value can be calculated by setting the bracket of the above equation to zero and solving for $\abs{A}^2$. As the phonon population is given by $\expval{n} = \expval{a^\dagger a} = A^*A= \abs{A}^2$ we have thus found the phonon expectation value above the lasing threshold which is given by
\begin{align}
    \expval{n}= \frac{\gamma_\mathrm{c} \gamma_\mathrm{h} \left(\gamma_\mathrm{c} g_\mathrm{h}^2 - \gamma_\mathrm{h} g_\mathrm{c}^2 \right)}{ 8g_\mathrm{c}^2 g_\mathrm{h}^2 (\gamma_\mathrm{c} - \gamma_\mathrm{h})}.
\end{align}
The second phase transition is more subtle. By analyzing Eq.~\eqref{eq:A:master} we note that each term corresponds to one of the ion species and shows a saturating behaviour: the larger $\abs{A}$, the smaller the contribution of each term becomes. As the two terms have opposite sign the question arises which term saturates faster. This is determined by the saturation coefficients $s_\mathrm{h/c}$ in front of $\abs{A(t)}^2$. For large values of $A$, this factor reduces to $1/\gamma_\mathrm{c/h}$ as we can neglect the $1$ in the denominator. In the case where $\gamma_\mathrm{c}>\gamma_\mathrm{h}$ the heating ion saturates before the cooling ion and therefore a finite steady state for $A$ can be reached. In the other case the heating ion saturates later and due to the positive sign of this term a runaway situation occurs which describes the heating phase. We conclude that the second phase transition to heating is given by the condition $\gamma_\mathrm{c}=\gamma_\mathrm{h}$ or equivalently by $s_\mathrm{h}/s_\mathrm{c} = \kappa_\mathrm{h}/\kappa_\mathrm{c}$. A phase diagram resulting from this analysis is shown in Fig.~\ref{fig:phaseDiagram:mf}.

If we include the dephasing operator explained and motivated in the main text into the description of the heating ion, the equations of motion for the heating spin have to be modified. \newline
The additional Lindblad jump operator, which implements the additional dephasing due to the 4-level nature of the system, is given by:
\begin{align}
    L_\mathrm{1}=\sqrt{\gamma_\mathrm{e}}\sigma_\mathrm{e}=\sqrt{\gamma_\mathrm{e}}\ket{1}\bra{1}.
\end{align}
This term modifies Eq.~\eqref{eq:mf:dgl:s} which now reads,
\begin{align}
    \frac{d}{dt}S_\mathrm{h}(t) &= -\frac{\gamma_\mathrm{h}}{2}S_\mathrm{h}(t)-\frac{\gamma_\mathrm{e}}{2}S_\mathrm{h}(t)-ig_\mathrm{h}A(t)D_\mathrm{h}(t).
\end{align}
The equations of motion of the other variables are not changed and we can follow the same procedure as above. By assuming the steady state for the spin variables, we find the evolution equation for the ion motion analogous to Eq.~\eqref{eq:A:master},
\begin{align}
    \frac{d}{dt}A(t) = A(t)\left(\frac{\frac{2g_\mathrm{h}^2}{(\gamma_\mathrm{h}+\gamma_\mathrm{e})}}{\frac{8 g_\mathrm{h}^2}{\gamma_\mathrm{h}(\gamma_\mathrm{h}+\gamma_\mathrm{e})} \abs{A(t)}^2 +1}-\frac{\frac{2g_\mathrm{c}^2}{\gamma_\mathrm{c}}}{\frac{8 g_\mathrm{c}^2}{\gamma_\mathrm{c}^2} \abs{A(t)}^2 +1}\right).
    \label{eq:A:master:mod}
\end{align}
By setting the brackets in the above equation to zero, we find the modified expression for the phonon occupation,
\begin{align}
    \expval{n}= \frac{\gamma_\mathrm{c} \gamma_\mathrm{h} \left(\gamma_\mathrm{c} g_\mathrm{h}^2 - (\gamma_\mathrm{h}+\gamma_\mathrm{e}) g_\mathrm{c}^2 \right)}{ 8g_\mathrm{c}^2 g_\mathrm{h}^2 (\gamma_\mathrm{c} - \gamma_\mathrm{h})}.
\end{align}
This expression is used in Fig.~\ref{fig:data1} (e) in the main text, to compare the measured and simulated phonon expectation value with the theory.

\section{Phase Diffusion: Heisenberg Langevin Approach}
\label{SM:phasediff_HL}
In order to model the decay of the heating ion as well as the oscillator decay (making use of the eliminated cooling ion), we have to introduce two reservoirs which couple to the heating ion and to the resonator respectively, as described in \cite{scully_zubairy}. The interactions between ion and cavity are described by the usual Anti-JC-Hamiltonian. \newline
The total Hamiltonian in the rotating wave approximation consists of the following three terms:
\begin{align}
        H_\mathrm{Ion;Motion} &= g\left(a^{\dagger}\sigma_\mathrm{+}+\sigma_\mathrm{-}a\right),\label{eq:hl:hamilton:ionmotion}\\
        H_\mathrm{Motion;Reservoir1} &= \sum_\mathrm{q} g_\mathrm{q}\left[c_\mathrm{q}^{\dagger}a e^{-i(\omega-\omega_\mathrm{q})t} + a^{\dagger}c_\mathrm{q} e^{i(\omega-\omega_\mathrm{q})t}\right],\label{eq:hl:hamilton:motionres1}\\
        H_\mathrm{Ion;Reservoir2} &= \sum_\mathrm{k} g_\mathrm{k}\left[\sigma_\mathrm{+}b_\mathrm{k} e^{i(\omega-\omega_\mathrm{k})t} + b_\mathrm{k}^{\dagger}\sigma_\mathrm{-} e^{-i(\omega-\omega_\mathrm{k})t}\right].\label{eq:hl:hamilton:ionres2}
\end{align}
The reservoirs consist of a collection of harmonic oscillators with closely spaced frequencies. $c_\mathrm{q}$ and $b_\mathrm{k}$ are bosonic destruction operators of these reservoir oscillators, which fulfill the commutation relations $[c_\mathrm{q},c_p^{\dagger}]=\delta_\mathrm{q,p}$ and $[b_\mathrm{k},b_l^{\dagger}]=\delta_\mathrm{k,l}$.\newline
We use the von Neumann equation to find the time evolution for the following operators:
\begin{align}
    \dot{a} &= -ig\sigma_\mathrm{+} -i\sum_\mathrm{q} g_\mathrm{q} c_\mathrm{q} e^{i(\omega-\omega_\mathrm{q})t},\label{eq:hl:eom:a}\\
    \dot{\sigma}_\mathrm{+} &= -ig\sigma_\mathrm{z}a -i\sum_\mathrm{k} g_\mathrm{k} b_\mathrm{k}^{\dagger} \sigma_\mathrm{z} e^{-i(\omega-\omega_\mathrm{k})t},\label{eq:hl:eom:sp}\\
    \dot{\sigma}_\mathrm{z} &= -2ig (a^{\dagger}\sigma_\mathrm{+} - \sigma_\mathrm{-}a) +2i\sum_\mathrm{k} g_\mathrm{k}\left[ -\sigma_\mathrm{+} b_\mathrm{k} e^{i(\omega-\omega_\mathrm{k})t} + b_\mathrm{k}^{\dagger}\sigma_\mathrm{-} e^{-i(\omega-\omega_\mathrm{k})t}\right],\label{eq:hl:eom:sz}\\
    \dot{c}_\mathrm{q} &= -ig_\mathrm{q} a e^{-i(\omega-\omega_\mathrm{q})t},\label{eq:hl:eom:c}\\
    \dot{b}_\mathrm{k} &= -ig_\mathrm{k}\sigma_\mathrm{-} e^{-i(\omega-\omega_\mathrm{k})t}.\label{eq:hl:eom:b}
\end{align}
Eqs.~\eqref{eq:hl:eom:c} and \eqref{eq:hl:eom:b} are formally integrated and the solution is substituted into Eq.~\eqref{eq:hl:eom:a} - \eqref{eq:hl:eom:sz}.\newline

From the resulting equations the noise operators and decay parameters $\gamma$ and $C$ can be identified and replaced by the following definitions,
\begin{align}
    \gamma &= 2\sum_\mathrm{k}g_\mathrm{k}^2\int_0^tdt'e^{-i(\omega-\omega_\mathrm{k})(t-t')},\label{eq:hl:def:gamma}\\
    C &= 2\sum_\mathrm{q}g_\mathrm{q}^2\int_0^tdt'e^{-i(\omega-\omega_\mathrm{q})(t'-t)},\label{eq:hl:def:c}\\
    \mathcal{F}_a(t) &= -i\sum_\mathrm{q} g_\mathrm{q} c_\mathrm{q}(0) e^{i(\omega-\omega_\mathrm{k})t},\label{eq:hl:def:fa}\\
    \mathcal{F}_\mathrm{+}(t) &= -i\sum_\mathrm{k} g_\mathrm{k} b_\mathrm{k}^{\dagger}(0)\sigma_\mathrm{z}(t) e^{-i(\omega-\omega_\mathrm{k})t},\label{eq:hl:def:fp}\\
    \mathcal{F}_\mathrm{z}(t) &= 2i\sum_\mathrm{k} g_\mathrm{k}\left[ -\sigma_\mathrm{+}(t) b_\mathrm{k}(0) e^{i(\omega-\omega_\mathrm{k})t} + b_\mathrm{k}^{\dagger}(0)\sigma_\mathrm{-}(t) e^{-i(\omega-\omega_\mathrm{k})t}\right].\label{eq:hl:def:fz}
\end{align}
With these definitions, Eq.~\eqref{eq:hl:eom:a}-\eqref{eq:hl:eom:sz} read:
\begin{align}
    \dot{a} &= -ig\sigma_\mathrm{+} -\frac{C}{2}a + \mathcal{F}_a,\label{eq:hl:eom:a:simple}\\
    \dot{\sigma}_\mathrm{+} &= -ig\sigma_\mathrm{z}a -\frac{\gamma}{2}\sigma_\mathrm{+} + \mathcal{F}_\mathrm{+},\label{eq:hl:eom:sp:simple}\\
    \dot{\sigma}_\mathrm{z} &= -2ig (a^{\dagger}\sigma_\mathrm{+} - \sigma_\mathrm{-}a) -\gamma (\sigma_z + 1) + \mathcal{F}_\mathrm{z}.\label{eq:hl:eom:sz:simple}
\end{align}
We can now calculate the correlation functions of the noise operators defined in Eq.~\eqref{eq:hl:def:fa} - \eqref{eq:hl:def:fz}. We assume that both reservoirs are in thermal equilibrium and that they are at zero temperature. This has the effect that $\Bar{n}_\mathrm{th}=0$ which makes most correlation functions zero. The non-vanishing ones are listed below:
\begin{align}
        \expval{\mathcal{F}_a(t)\mathcal{F}_a^{\dagger}(t')}_\mathrm{R}&=C\delta(t-t'),\label{eq:hl:correlator:aa}\\
        \expval{\mathcal{F}_+^{\dagger}(t)\mathcal{F}_+(t')}_\mathrm{R}&=\gamma\delta(t-t'),\label{eq:hl:correlator:pp}\\
        \expval{\mathcal{F}_z(t)\mathcal{F}_z^{\dagger}(t')}_\mathrm{R}&=2\gamma\left[1+\expval{\sigma_z(t)}\right]\delta(t-t'),\label{eq:hl:correlator:zz}\\
        \expval{\mathcal{F}_+^{\dagger}(t)\mathcal{F}_z(t')}_\mathrm{R}&=2\gamma\expval{\sigma_-(t)}\delta(t-t'),\label{eq:hl:correlator:pz}\\
        \expval{\mathcal{F}_z(t)\mathcal{F}_+(t')}_\mathrm{R}&=2\gamma\expval{\sigma_+(t)}\delta(t-t').\label{eq:hl:correlator:zp}
\end{align}
Note that $\mathcal{F}_z$ is a Hermitian operator, therefore also all correlations functions above with $\mathcal{F}_z$ replaced by its adjoint result in the same function.\newline
In order to solve the operator Eq.~\eqref{eq:hl:eom:a:simple} - \eqref{eq:hl:eom:sz:simple} we have to find the corresponding complex number equations. For this we fix a normal ordering of the operators: $a^{\dagger},\sigma_\mathrm{-},\sigma_\mathrm{z},\sigma_\mathrm{+},a$. Any operator equation which is written in this defined order can be translated into a complex number equation. As the Eq.~\eqref{eq:hl:eom:a:simple} - \eqref{eq:hl:eom:sz:simple} are already in this order, all operators can be replaced with the complex variables $A,S,D$ and we obtain the following equations
\begin{align}
    \dot{A} &= -igS - \frac{C}{2}A + F_a, \label{eq:hl:eom:a:complex}\\
    \dot{S} &= -igDA - \frac{\gamma}{2} S + F_+, \label{eq:hl:eom:s:complex}\\
    \dot{D} &= -2ig(A^{*}S - S^{*}A)  - \gamma(D + 1) + F_z.\label{eq:hl:eom:d:complex}
\end{align}
The noise operators $F_a,F_+,F_z$, which are complex valued functions, can be determined by requiring that the time evolution of the second order moments of the complex variables are the same as those of the operator variables. Note this requirement guarantees that the commutation relations of the operators remain time invariant.\newline

We use the following definition of the diffusion coefficients to shorten the notation of the noise correlator functions:
\begin{equation}
    \expval{F_\mu(t)F_\nu(t')}=\expval{2D_\mathrm{\mu\nu}}\delta(t-t').
\end{equation}
The nonzero diffusion coefficient yield:
\begin{align}
        \expval{2D_\mathrm{++}}&=2ig\expval{\sigma_+a}\label{eq:hl:diffcoeff:pp}\\
        \expval{2D_\mathrm{zz}}&=+4ig\expval{a^{\dagger}\sigma_+}-4ig\expval{\sigma_-a}+2\gamma(1+\expval{\sigma_z})\label{eq:hl:diffcoeff:zz}\\
        \expval{2D_\mathrm{--}}&=-2ig\expval{a^{\dagger}\sigma_-}\label{eq:hl:diffcoeff:mm}\\
        \expval{2D_\mathrm{-z}}&=2\gamma\expval{\sigma_-}\label{eq:hl:diffcoeff:mz}\\
        \expval{2D_\mathrm{z+}}&=2\gamma\expval{\sigma_+}\label{eq:hl:diffcoeff:zp}\\
        \expval{2D_\mathrm{-+}}&=\gamma\label{eq:hl:diffcoeff:mp}
\end{align}
In order to solve the Eq.~\eqref{eq:hl:eom:a:complex} - \eqref{eq:hl:eom:d:complex}, the adiabatic approximation is made, which allows us to assume the steady state for the spin variables $S$ and $D$. This gives us one equation for the evolution of $A$ with its noise operator $F_\mathrm{a}$:

\begin{equation}
\begin{aligned}
    &\dot{A} = -\frac{2g^2}{\gamma}\frac{1}{\gamma(1+\frac{8g^2}{\gamma^2}I)}\left[-\frac{4ig}{\gamma} (A^{*}F_\mathrm{+} -F_\mathrm{+}^{*}A) - \gamma + F_z\right]A + -\frac{2ig}{\gamma}F_\mathrm{+} - \frac{C}{2}A + F_a\\
    &=\frac{\frac{2g^2}{\gamma}A}{(1+\frac{8g^2}{\gamma^2}I)} -\frac{C}{2}A + F_\mathrm{A},
\end{aligned}
\end{equation}
with
\begin{equation}
    F_\mathrm{A} =  \frac{1 }{(1+\frac{8g^2}{\gamma^2}I)}\left[\frac{8ig^3}{\gamma^3}(A^{*}F_\mathrm{+} -F_\mathrm{+}^{*}A)- \frac{2g^2}{\gamma^2}F_\mathrm{z}\right]A+\frac{2g}{\gamma}F_\mathrm{+} + F_\mathrm{a}.
\end{equation}
This new total noise operator $F_A$ now contains all information about the noise induced to $A$. In order to use this result further, we have to calculate the diffusion coefficient for it (i.e., the correlator functions between combinations of $F_A$ and $F_\mathrm{A}^{*}$)
\begin{equation}
    \begin{aligned}
        \expval{2D_\mathrm{AA}}&=\expval{F_A(t)F_A(t)}=\frac{-16A^2}{(1+\frac{8g^2}{\gamma^2}I)^3}\left[32\frac{g^8}{\gamma^7}I^2+16\frac{g^6}{\gamma^5}I+3\frac{g^4}{\gamma^3}\right],
    \end{aligned}
\end{equation}
\begin{equation}
    \begin{aligned}
        \expval{2D_\mathrm{A^{*}A}}&=\expval{F_A^{*}(t)F_A(t)}=\frac{4}{1+(\frac{8g_\mathrm{h}^2}{\gamma_\mathrm{h}^2}I)^3}\left[128\frac{g_\mathrm{h}^8}{\gamma_\mathrm{h}^7}I^3+64\frac{g_\mathrm{h}^6}{\gamma_\mathrm{h}^5}I^2+8\frac{g_\mathrm{h}^4}{\gamma_\mathrm{h}^3}I+\frac{g^2}{\gamma}\right].
    \end{aligned}
\end{equation}

Now in order to get the phase diffusion coefficient we make the following coordinate transformation: $A=\sqrt{I}e^{i\theta}$. From the identity
\begin{equation}
\frac{\dot{A}}{A}=\frac{\dot{I}}{2I}+i\dot{\Theta},
\end{equation}
we can see that the purely imaginary part of $\frac{\dot{A}}{A}$ describes the evolution in tangential direction and the real part correspondingly the radial evolution of the state.\newline
In a similar way the noise operator can be split into real and imaginary part where the imaginary part corresponds to the phase noise. In order to find the diffusion coefficient for the phase noise which is given by $\expval{2D_\mathrm{\Theta\Theta}}=\expval{\operatorname{Im}(\frac{F_A(t)}{A})\operatorname{Im}(\frac{F_A(t)}{A})}$, we can use the following identity for an arbitrary complex number $x=a+ib$.
\begin{equation}
    4\operatorname{Im}(x)^2=xx^{*}+x^{*}x-x^{*}x^{*}-xx 
\end{equation}
By setting $x=\frac{F_A(t)}{A}$ we find
\begin{equation}
    4\expval{2D_\mathrm{\Theta\Theta}}=\frac{\expval{2D_\mathrm{AA^{*}}}}{I}+\frac{\expval{2D_\mathrm{A^{*}A}}}{I}-\frac{\expval{2D_\mathrm{AA}}}{A^2}-\frac{\expval{2D_\mathrm{A^{*}A^{*}}}}{(A^{*})^2}.
\end{equation}
Using the diffusion coefficients of the complex noise operators $F_\mathrm{+},F_\mathrm{-}$ and $F_\mathrm{z}$ as given in Eqs.~\eqref{eq:hl:diffcoeff:pp}-\eqref{eq:hl:diffcoeff:mp}. We can find the following explicit expression of the phase diffusion coefficient in terms of the system parameters:
\begin{equation}
    \expval{2D_\mathrm{\Theta\Theta}}=\frac{\frac{2g^2}{\gamma}+\frac{8g^4I}{\gamma^3}}{I(1+\frac{8g^2}{\gamma^2}I)}.
    \label{eq:diffusionnogammae}
\end{equation}

\section{Phase Diffusion: Heisenberg Langevin Approach including dephasing operator \texorpdfstring{$\sigma_\mathrm{e}$}{sigmae}} 
\label{SM:phasediff_HL_wdephasing}
We observed a mismatch in the average phonon number as well as the phase diffusion between experiment and simulations considering beryllium as a two-level system. The mismatch is reduced with simulations of the full four level beryllium system. We find that we need to consider a projective operator $\sigma_\mathrm{e}=\ket{1}\bra{1}$ at rate $\gamma_\mathrm{e}=f\gamma_\mathrm{h}$ (where $f=50/40$ representing the branching ratios) to approximate our beryllium system with a two-level system, including the effect of the decay from the P-state back into $\ket{1}$. In doing so, we get agreement of experiment and simulations in the mean phonon number. We therefore decide to add this term to the Heisenberg Langevin approach for the phase diffusion. The new set of Hamiltonians now contains an additional reservoir and is given by
\begin{align}
        H_\mathrm{Ion;Motion} &= g\left(a^{\dagger}\sigma_\mathrm{+}+\sigma_\mathrm{-}a\right),\\
        H_\mathrm{Motion;Reservoir1} &= \sum_\mathrm{q} g_\mathrm{q}\left[c_\mathrm{q}^{\dagger}a e^{-i(\omega-\omega_\mathrm{q})t} + a^{\dagger}c_\mathrm{q} e^{i(\omega-\omega_\mathrm{q})t}\right],\\
        H_\mathrm{Ion;Reservoir2} &= \sum_\mathrm{k} g_\mathrm{k}\left[\sigma_\mathrm{+}b_\mathrm{k} e^{i(\omega-\omega_\mathrm{k})t} + b_\mathrm{k}^{\dagger}\sigma_\mathrm{-} e^{-i(\omega-\omega_\mathrm{k})t}\right],\\
        H_\mathrm{\ket{1}\bra{1};Reservoir3}&= \sum_\mathrm{l} g_\mathrm{l}\left[\sigma_\mathrm{e}^{\dagger}d_\mathrm{l} e^{i(\omega-\omega_\mathrm{l})t} + d_\mathrm{l}^{\dagger}\sigma_\mathrm{e} e^{-i(\omega-\omega_\mathrm{l})t}\right].
\end{align}
The derivation of the diffusion coefficient is analogous to the previous chapter. The nonzero diffusion coefficients are:
\begin{align}
        \expval{2D_\mathrm{++}}&=2ig\expval{\sigma_+a},\\
        \expval{2D_\mathrm{zz}}&=+4ig\expval{a^{\dagger}\sigma_+}-4ig\expval{\sigma_-a}+2\gamma(1+\expval{\sigma_z}),\\
        \expval{2D_\mathrm{--}}&=-2ig\expval{a^{\dagger}\sigma_-},\\
        \expval{2D_\mathrm{-z}}&=2\gamma\expval{\sigma_-},\\
        \expval{2D_\mathrm{z+}}&=2\gamma\expval{\sigma_+},\\
        \expval{2D_\mathrm{-+}}&=\gamma+\frac{\gamma_\mathrm{e}}{2}(\langle\sigma_z\rangle-1).
\end{align}

Following a similar approach to Sec.~\ref{SM:phasediff_HL}, we use these to find the modified phase diffusion which we use to evaluate the contribution from the beryllium ion
\begin{equation}
    \expval{2D_\mathrm{\Theta\Theta}}=\frac{\frac{2g^2}{(\gamma+\gamma_\mathrm{e})}+\frac{8g^4I}{(\gamma (\gamma+\gamma_\mathrm{e})^2)}}{I\left(1+\frac{8g^2}{(\gamma(\gamma+\gamma_\mathrm{e}))}I\right)}.
\end{equation}
This reduces to the expression in Eq.~\eqref{eq:diffusionnogammae} in the limit that $\gamma_\mathrm{e} \rightarrow 0$.